\documentclass[letter,twocolumn,showpacs,aps,prb,floatfix,superscriptaddress]{revtex4}
\usepackage{graphicx,subfigure,epsfig,verbatim,psfrag,amsmath,amssymb,color}

\makeatletter
\input epsf

\def\be{\begin{equation}}
\def\ee{\end{equation}}
\def\ba{\begin{eqnarray}}
\def\ea{\end{eqnarray}}
\makeatother

\begin{document}

\title{Criticality and Mott-glass phase in a disordered 2D quantum spin systems}

\author{Nvsen Ma}
\affiliation{State Key Laboratory of Optoelectronic Materials and
Technologies, School of Physics and Engineering, Sun Yat-Sen
University, Guangzhou 510275, China}

\author{Anders W. Sandvik}
\email[e-mail:]{sandvik@bu.edu}
\affiliation{Department of Physics, Boston University, 590 Commonwealth Avenue, Boston, MA 02215, USA}

\author{Dao-Xin Yao}
\email[e-mail:]{yaodaox@mail.sysu.edu.cn}
\affiliation{State Key Laboratory of Optoelectronic Materials and
Technologies, School of Physics and Engineering, Sun Yat-Sen
University, Guangzhou 510275, China}

\date{\today}

\begin{abstract}
We use quantum Monte Carlo simulations to study a disordered $S=1/2$ Heisenberg quantum spin model with three different nearest-neighbor interactions,
$J_1 \le J_2 \le J_3$, on the square lattice. We consider the regime in which $J_{1}$ represents weak bonds, and $J_{2}$ and $J_{3}$ correspond to two
kinds of stronger bonds (dimers) which are randomly distributed on columns forming coupled 2-leg ladders. When increasing the average intra-dimer coupling
$(J_{2}+J_{3})/2$, the system undergoes a N$\acute{e}$el to quantum glass transition of the ground state and later a second transition into a quantum paramagnet.
The quantum glass phase is of the gapless Mott glass type (i.e., in boson language it is incompressible at temperature $T=0$), and we find that the temperature
dependence of the uniform magnetic susceptibility follows the stretched exponential form $\chi\sim\exp(-b/T^{\alpha})$, with $0<\alpha<1$. At the N\'eel--glass
transition we observe the standard O(3) critical exponents, which implies that the Harris criterion for the relevance of the disorder is violated in this system.

\end{abstract}
 \maketitle

\section{Introduction}
\label{sec:intro}

When some form of disorder is introduced in a quantum many-body system,
e.g., random coupling constants, impurities, or dilution of the degrees of freedom, the interplay between quantum fluctuations
and disorder effects can lead to unconventional properties and special quantum phases.~\cite{overall} Such phenomena
have been the subject of enormous interest in both theoretical and experimental physics, e.g., Anderson localization of non-interacting
electrons,\cite{lee85} anomalous metal-insulator transition in interacting two-dimensional electron systems,\cite{abrahams01} Bose-Einstein condensation in
disordered boson systems,\cite{fisher,BEC} which can be experimentally realized, e.g., in $^{4}$He films on substrates with disorder \cite{hefour} and in quantum magnets
with random couplings.\cite{fisher94,bbec,nature12} Disorder also likely plays an important role in many strongly-correlated systems with superconducting
phases.\cite{superconductor} Going beyond ground-state properties, the low-energy excitations can also change completely in the presence of disorder,
\cite{wangperc,changlani13} and many-body localization of higher excited states \cite{basko06} is now also attracting considerable attention. In this paper we
study the effects of disorder on an equilibrium quantum phase transition \cite{sachdev} between a magnetic and quantum paramagnetic state in a two-dimensional (2D)
quantum spin model with tunable couplings. We also discuss related ``quantum glass'' physics.

The standard 2D Heisenberg model, which serves as the foundation of our understanding of the physics of the insulating state of the high-Tc cuprates
and many other quantum antiferromagnets,\cite{manousakis91,sandvik10}  has been the subject of numerous theoretical and computational studies of disorder
effects.\cite{chen00,kato00,sandvik01,sandvik02,yu05,laflorencie06,liu06}  Experimental realizations also have been investigated extensively.\cite{vajk02,carretta11}
We here study ground-state and finite-temperature properties of a class of disordered 2D Heisenberg models with the goal of establishing some key
bench-mark results for quantum criticality and associated glass phases.

This work has a broader context of disorder in boson systems (to which spin
models can be mapped), where randomness in the local potentials and couplings can destroy superfluid long-rang order and bring about different types of glass
phases~\cite{fisher,glassphase} which are distinct from the corresponding quantum-disordered phases in clean
systems. Two classes of glass phases have been identified---the Bose glass, which is compressible at temperature $T =0$, and the incompressible
Mott glass (MG).\cite{giamarchi01} The latter is in general believed to exist only in systems with particle-hole symmetry,\cite{boson1d,glassphase} though recent
computational work has found behaviors typical for MGs also in the absence of this symmetry \cite{wang11} and some theoretical
arguments also support its possible existence more generally.\cite{giamarchi01}

Based on a strong-disordered renormalization group method applied to an effective model representing the one-dimensional (1D) Bose-Hubbard model at large
integer filling fraction and local particle-hole symmetry, Altman {\it et al.} found that the transition into the MG state is in the Kosterliz-Thouless
universality class with a dynamical exponent $z=1$.~\cite{boson1d}  Iyer \emph{et al.} extended the calculations to two dimensions and presented evidence for the
existence of the MG phase also in this case.\cite{mglass03} This result agrees with quantum Monte Carlo (QMC) simulations of the 2D Bose-Hubbard model
with random hopping (which maintains local particle-hole symmetry), where a non-trivial dynamic exponent $z \approx 1.5$ was found.~\cite{glassphase} This special
incompressible glass phase was also recently claimed to be experimentally realized in a doped quantum magnet (Br-DTN).~\cite{nature12}

Although there are many computational works identifying intriguing quantum-critical phenomena and ground state phases in various disordered $S=1/2$ Heisenberg quantum spin
models,\cite{chen00,kato00,sandvik01,sandvik02,yu05,laflorencie06,liu06,wangperc,changlani13,yao10} so far the MG phase has not been the focus on such studies, except
for a diluted spin-1 model investigated in Ref.~\onlinecite{roscilde07}. Based on general arguments for the existence of Griffiths phases adjacent to critical points
in disordered systems,\cite{sachdev} one should also expect MG phases in spin-isotropic $S=1/2$ spin system (which have, in boson language, particle-hole symmetry)
with critical points when randomness is introduced in the couplings in some way. We here present a detailed study of a class of disordered 2D dimerized $S=1/2$
quantum spin models, confirming the existence of an MG phase and studying its properties both at $T=0$ and $T>0$.

In addition to uncovering MG physics, another key aspect of this work is to study the influence of disorder on the quantum-critical scaling behavior. In a pioneering
study of the effects of disorder at phase transitions, Harris derived a criterion for the critical exponents based on a classical Ising model.\cite{harris} According
to the Harris criterion, the critical behavior (universality class) will not be influenced by disorder if the correlation length exponent $\nu$ satisfies
$\nu>2/d$, where $d$ is the dimensionality of the system. If the inequality does not hold, the universality class should change upon the introduction of disorder,
so that $\nu>2/d$ applies (i.e., the disorder is a relevant perturbation). The Harris criterion was later rederived under more general conditions,\cite{chayes86}
applying also to a wide range of quantum systems. For a quantum system it is believed that the dimensionality $d$ to use in the inequality should be the spatial
dimensionality $d$ (if the disorder is introduced symmetrically in all the spatial dimensions), i.e., not the space-time dimensionality $d+1$ representing the effective
statistical-mechanical system obtained in a path-integral representation at $T=0$.\cite{chayes86,sachdev} The effective system has columnar disorder with constant
couplings along the time dimension.

Many studies of classical systems with disorder have been in agreement with the Harris criterion. In a work of relevance to 2D Heisenberg quantum spins,
a $(2+1)$-dimensional Heisenberg model with columnar dilution was studied and it was found that  $\nu \approx 1.1 - 1.2 >2/d=1$,\cite{vajk02,fulfill} while in the clean
system $\nu \approx 0.71$ of the 3D O(3) universality class. However in a class of 2D $S=1/2$ dimerized quantum spin models with random arrangements of weaker
(intra-dimer) and stronger (inter-dimer) interactions,\cite{yao10} no change in the exponents with respect to to the clean 3D O(3) exponents were found, and the
Harris criterion is then violated. Possible violations of the Harris criterion in a disordered boson system were discussed in Refs.~\onlinecite{kisker97,pazmandi98}.
The general validity of the Harris criterion has also been questioned theoretically,\cite{pazmandi97} and as of now there does not appear to be any rigorous way to
establish {\it a priori} (without elaborate explicit calculations of critical exponents) whether a particular kind of disorder is relevant or not.

\begin{figure}
\centering
{\resizebox*{0.2\textwidth}{!}{\includegraphics{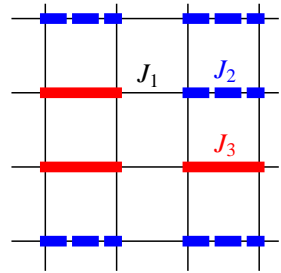}}}
\caption{(Color online). 2D square-lattice columnar-dimerized Heisenberg model with two types of strong bonds. Blue and red thick lines represent the strong bonds,
which reside on vertical columns separated by one lattice spacing. The corresponding couplings, $J_{2}$ or $J_{3}$, are chosen at random. The thin lines stand
for the weaker inter-dimer coupling $J_{1}=1$.}
\label{modelf}
\end{figure}

Our studies of criticality and MG physics reported in this paper were carried out within a dimerized $S=1/2$ Heisenberg model starting from the well-studied columnar dimer
pattern.\cite{singh88,matsumoto01,wenzel09,sandvik10} In a clean system with inter- and intra-dimer couplings $J_1$ and $J_2$, the N\'eel order of the ground state existing for
or $J_2 \approx J_1$ vanishes when $J_2/J_1\approx 1.9$ and the excitation spectrum becomes gapped. Universality of this transition in the O(3) class has been confirmed
to numerical precision rivaling that in studies of classical models.\cite{matsumoto01,wenzel09} We here introduce randomness in the intra-dimer couplings, using a total
of three different antiferromagnetic nearest-neighbor interactions, $J_{1}$, $J_{2}$ and $J_{3}$, with $J_1 \le J_2 \le J_3$ and setting the inter-dimer coupling $J_1=1$. The
stronger, intra-dimer couplings again form columns, with random placements of $J_2$ and $J_3$ bonds within these columns. Fig.~\ref{modelf} shows an illustration of the system.
Using the Stochastic Series Expansion (SSE) QMC method,\cite{sse} we find an MG phase with magnetic susceptibility vanishing as $T \to 0$, following the form
$\chi\sim\exp(-b/T^{\alpha})$, with $\alpha$ depending on the parameters (and $\alpha =1$ in the standard gapped phase as in the clean system). Using various
quantities for finite-size scaling we find no detectable differences between the clean and the random system. The correlation length exponent $\nu \approx 0.71 < 2/d$,
and the Harris criterion is therefore violated.

This rest of the paper is organized as follows. In Sec.~\ref{sec:model} we define the random columnar spin model in detail and discuss the physical observables
studied and how we approach the $T\to 0$ limit in the QMC simulations. In Sec.~\ref{sec:critical} we present finite-size scaling studies to extract the critical
point and the critical exponents. The susceptibility in the MG phase is discussed in Sec.~\ref{sec:mg}, followed by a brief summary and discussion in
Sec.~\ref{sec:disc}.

\section{Model, observables, methods}
\label{sec:model}

\subsection{Hamiltonian}

The Hamiltonian of the  disordered antiferromagnetic columnar-dimerized Heisenberg model can be simply written as
\begin{equation}
   H=\sum_{\langle ij\rangle}J_{ij} {\bf S}_{i}\cdot {\bf S}_{j},
\label{hamiltonian}
\end{equation}
where ${\bf S}_{i}$ is an $S=1/2$ spin operator residing on an $L\times L$ square lattice with even $L$ and periodic boundary conditions. The index pairs $\langle ij\rangle$
denote nearest-neighbor sites with corresponding coupling strengths $J_{ij} \in \{ J_1,J_2,J_3\}$. As illustrated in Fig.~\ref{modelf}, the bonds where
$J_{ij} \in \{J_2,J_3\}$ form columns separated by one bond and the assignment to $J_2$ or $J_3$ is done at random with probability $1/2$ for each choice.
All other couplings are assigned the inter-dimer value $J_1=1$.

In order to more conveniently describe the intra-dimer couplings in terms of an overall strength and a fluctuation, we parametrize them according to
  \begin{align}
  J_{2} &= 1 + (g - 1)(1 - p), \\
  J_{3} &= 1 + (g - 1)(1 + p),
  \end{align}
so that the average is $(J_{2} + J_{3})/2 = g$ and the difference $J_3-J_2=2p(g-1)$. Normally we chose $g>1$ and the parameter $0\leq p \leq 1$ serves as a measure
of the strength of the disorder.

In the clean limit $p=0$, when $g=1$ our model reduces to the isotropic Heisenberg model, which is N\'eel-ordered at $T=0$ (but not at $T>0$ by the
Mermin-Wagner theorem), while for $g \to \infty$ the ground state is a product of dimer singlets and there is no magnetic order. Several QMC studies have been devoted
to the quantum phase transition between these two limiting cases,\cite{singh88,matsumoto01,wenzel09,sandvik10} which takes place at $g_{c}=1.90948(4)$ according
to the most precise calculation.\cite{sandvik10} In this paper we focus on a fixed rather strong disorder strength, $p=1/2$, where at the critical point the strong to weak
dimer ratio is $J_3/J_2 \approx 1.6$.  We use the SSE QMC method to compute several physical quantities averaged over $100-1000$ realizations of the random bonds in
each case.

\subsection{Observables}

We here define the physical quantities studied in this work.
The spin stiffness $\rho_{s}$ characterizes the response to twisting the magnetic order, similar to an elastic constant.
It is defined as
\begin{equation}
  \rho_{s}\equiv\frac{1}{N}\frac{\partial^{2}F(\phi)}{\partial\phi^{2}},
\label{spinstiffness}
\end{equation}
where $F$ is the free energy of the system and $\phi$ is the angle of a twist imposed between two column of spins (in the $x$ or $y$ direction of the lattice). The
stiffness is evaluated in SSE simulations using winding number fluctuations.\cite{sse} Another important quantity characterizing the magnetic state is the static
spin structure factor $S(\mathbf{q})$, i.e., the Fourier transform of the spin-spin correlation function at wave-vector $\mathbf{q}$
\begin{equation}
       S(\mathbf{q}) = \sum_{\mathbf{r}}e^{-i\mathbf{q}\cdot\mathbf{r}}C(\mathbf{r})=\sum_{\mathbf{r}}\cos(\mathbf{q}\cdot\mathbf{r})C(\mathbf{r}),
\label{strufac}
\end{equation}
where $C(\mathbf{r})$ is the correlation function
\begin{equation}
      C(\mathbf{r}_{i}-\mathbf{r}_{j})=\langle S_{i}^{z}S_{j}^{z} \rangle=\frac{1}{3}\langle {\bf S}_{i}\cdot {\bf S}{j} \rangle,
\label{corrfuc}
\end{equation}
which is a function only of the separation between the two spins after disorder-averaging.
Further, we study the magnetic susceptibility in real space, i.e., the linear response of a spin $j$ to a field coupled to spin $i$;
\begin{equation}
\chi(i,j) = \int^{\beta}_{0}d\tau\langle S^{z}_{i}(0)S^{z}_{j}(\tau)\rangle,
\label{staggeredx}
\end{equation}
where $S^{z}_{j}(\tau) = {\rm e}^{-\tau H}S^z_j {\rm e}^{\tau H}$. Again, after disorder averaging this susceptibility depends only on the separation
${\bf r}_{ij}$ of the two spins and we again carry out the Fourier transform to wave-vector ${\bf q}$. Both the structure factor and the susceptibility have simple
SSE estimators, in the case of the susceptibility based on computing the integration over $\tau$ exactly.\cite{sse} We here present results at the ordering
wave-vector $\mathbf{q}=(\pi,\pi)$, and, in the case of the susceptibility also at $q=0$ where the response to a uniform field reduces to
\begin{equation}
      \chi_{u}=\chi(0,0)=\frac{\beta}{N}\left \langle \left (\sum^{N}_{i=1}S^{z}_{i}\right )^{2}\right \rangle,
\label{unifomx}
\end{equation}
because of the Hamiltonian conserving the total magnetization.

\subsection{Convergence properties}
\label{sec:conv}

\begin{figure}
\centering
{\resizebox*{0.7\textwidth}{!}{\includegraphics{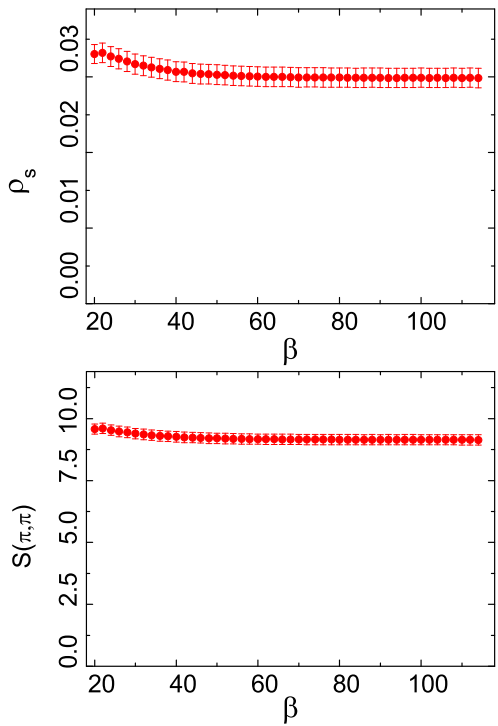}}}
\caption{(Color online) The spin stiffness (upper panel) and the staggered structure factor (lower panel) versus the inverse temperature for $L=32$ systems
at coupling ratio $g=1.98$, $p=1/2$ (close to the quantum critical point). Both of the quantities exhibit sufficient convergence within statistical
errors when $\beta \agt 60$.}
\label{converg1}
\end{figure}

\begin{figure}
\centering
{\resizebox*{0.7\textwidth}{!}{\includegraphics{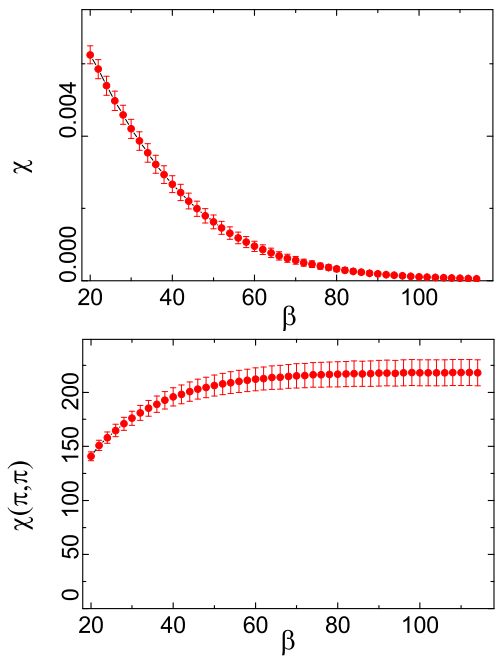}}}
\caption{Color online) The staggered (lower panel) and uniform (upper panel) susceptibilities versus the inverse temperature $\beta$ for $L=32$ systems at $g=1.98$,
$p=1/2$. The uniform susceptibility tends to $0$ when $\beta \to \infty$ and is not useful for $T=0$ studies, while the staggered susceptibility converges to a
finite non-zero value.
\label{converg2}}
\end{figure}

In order to use finite-temperature QMC simulations to investigate the ground state, it is necessary to make sure that the $T=0$ convergence is achieved
or, when studying a continuous quantum phase transition, to scale the inverse temperature $\beta=J_1/T$ with the system size using the proper dynamic
exponent; $\beta \propto L^z$. In this case we do not {\it a priori} know the value of $z$ and it is therefore better to converge to the true ground state of the system.
Figs.~\ref{converg1} and \ref{converg2} show examples of the convergence with
increasing $\beta$ of the quantities we discussed above for systems with $L=32$. Here the error bars primarily reflect sample-to-sample fluctuations affecting
the disorder averages. The same disorder realizations were used for all the temperatures, which implies that the statistical errors at different temperatures
are strongly correlated (again because the errors are dominated by sample-to-sample fluctuations, not the statistical errors of the individual simulations),
which is clearly visible in the data in Figs.~\ref{converg1} and \ref{converg2}.

The ground state of the model always being a singlet, the uniform susceptibility vanishes as $T\to 0$,
and we will only analyze its $T>0$ behavior. For the other quantities we find that $\beta=2L$ is sufficient
(for the system sizes considered here) to achieve reasonable convergence to the ground state, in the sense that remaining temperature effects are much
smaller than the statistical errors in the disorder averaging. This scaling of $\beta$ is also of course suitable for studying $z=1$ criticality (which
is in the end what we will find for the dynamic exponent), regardless of any remaining finite-temperature effects.

As eluded to above, even for relatively short SSE simulations of individual disorder realizations, the final statistical averages are dominated by the sample-to-sample
fluctuations, not the fluctuations of the individual SSE averages. It is therefore desirable to perform short simulations of a large number of samples. In order to
avoid potential problems of poorly equilibrated simulations it is then useful to use the {\it $\beta$-doubling procedure},\cite{sandvik02} where an almost equilibrated
configuration to start from in a simulation at inverse temperature $\beta$ is obtained from a prior simulation at $\beta'=\beta/2$ by combining two copies of the last sampled
configuration into a single configuration with twice the extent in the imaginary time direction.  For discussion of this approach and typical equilibration times
we refer to Ref.~\onlinecite{sandvik02}.

We will also discuss a set of calculations aiming at reaching the thermodynamic limit for $T>0$. In this case, as $T$ is lowered, increasingly
large system sizes have to be used in order to achieve infinite-size convergence. This convergence cannot easily be achieved for divergent quantities
(at criticality and in the N\'eel phase) at low $T$, because $L$ has to exceed the exponentially divergent \cite{nelson} correlation length. However, for
non-divergent quantities such as the uniform susceptibility convergence can be achieved at relatively low temperatures, as we will discuss later.

\section{Antiferromagnetic quantum phase transition}
\label{sec:critical}

In this section we use results of large-scale SSE studies with the $\beta$-doubling scheme to study critical properties of the ground state
of the system at disorder strength $p=1/2$. We use several finite-size scaling approaches to locate the critical coupling ratio $g$ where the magnetic
long-range order vanishes and extract the dynamical exponent $z$ along with with the equilibrium critical exponents $\nu$ and $\eta$.
We also study the magnetic susceptibility as a function of temperature in the thermodynamic limit.

\subsection{Critical point and dynamical exponent}

A common way to locate quantum phase transitions by finite-size scaling is to examine a quantity which is expected to be independent
of the system size at the critical point. A useful quantity for analyzing the loss of magnetic order in a spin systems is the spin
stiffness, Eq.~(\ref{spinstiffness}), which close to criticality should obey the scaling form
\begin{equation}
      \rho_s(L,g)=L^{-z}f[(g-g_{c})L^{{1}/{\nu}}],
\label{fsscform}
\end{equation}
where $\nu$ is the correlation length exponent (see Ref.~\onlinecite{sandvik10} for a review of the above form and other scaling behaviors of spin systems discussed below).
This form holds for the ground state, $\beta \to \infty$, and also for low but non-zero temperatures if $\beta$ is scaled proportionally to $L^z$. The properly
size-scaled stiffness $\rho L^{z}$ should then be independent of $L$ when $g=g_{c}$, provided that the system sizes are sufficiently large for corrections to scaling to
be negligible. If $z$ is known, one can then graph $\rho L^z$ versus $g$ for several system sizes and examine crossing points of curves for different $L$. In the
presence of scaling corrections (which cannot be avoided in practice) these crossing points will only tend toward the critical point when $L\to \infty$.

\begin{figure}
\centering
{\resizebox*{0.5\textwidth}{!}{\includegraphics{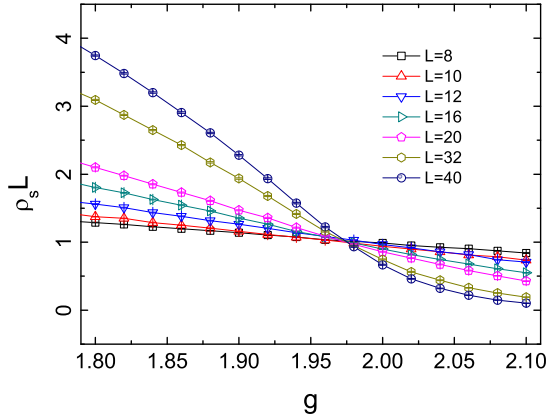}}}
\caption{(Color online). The size-scaled spin stiffness, assuming a dynamic exponent $z=1$, versus the average dimer coupling $g$ for different system sizes.
The inverse temperature is scaled with the size as $\beta=2L$ and the results were averaged over $1000$ disorder realizations. Curves for different $L$ are expected
to cross each other at the critical coupling ratio.
\label{cross}}
\end{figure}

For our system at hand here, we do not {\it a priori} know the value of $z$, but it should be noted that even if the wrong value is used there will
still be curve crossings and these will still tend toward the correct critical point when $L \to \infty$. However, the points will drift in the vertical direction
instead of converging to a stationary point. For now we use $z=1$ scaling with SSE data obtained at inverse temperature $\beta=2L$, which, as we discussed in
Sec.~\ref{sec:conv}, produces results essentially converged to the ground state for the system sizes and $g$ values we are considering. With practically converged
ground state results we are effectively in the $\beta \to \infty$ regime and can also test scaling using any value of $z$ (i.e., without scaling $\beta \propto L^z$ in
different simulations for each $z$ tested). In Fig.~\ref{cross} we use $z=1$ and plot disorder-averaged results for $\rho L$ versus $g$ for different $L$ and
$g \in [1.8,2.1]$. The crossing point is well defined even for rather small systems, with only minor drifts that we will examine in detail below.

\begin{figure}
\centering
{\resizebox*{0.55\textwidth}{!}{\includegraphics{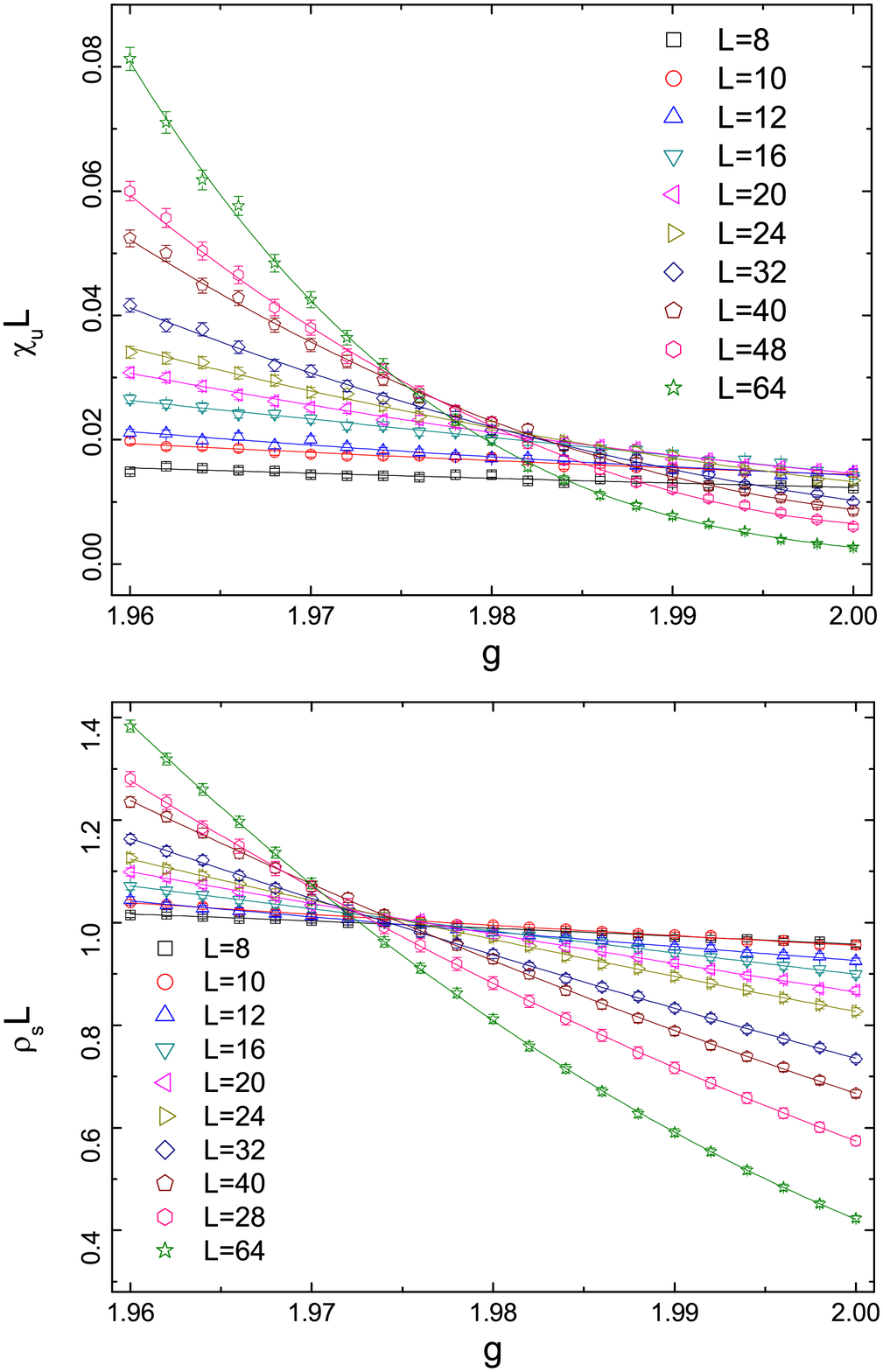}}}
\caption{(Color online). Size-scaled uniform susceptibility (upper panel) and spin stiffness (lower panel) with dynamic exponent $z=1$ assumed.
The data for each $L$ were averaged over $1000$ configurations. The curves show fitted polynomials, using which crossing points are extracted.
The horizontal and vertical drifts of the crossing points are analyzed in Figs.~\ref{criticalpoint} and \ref{criticaly}, respectively.
\label{critical}}
\end{figure}

The uniform susceptibility $\chi_{u}=\chi(0,0)$ is also a very useful quantity to consider in finite-size scaling. At the critical point we expect \cite{fisher}
\begin{equation}
\chi_{u}\sim L^{z-2}
\label{xuscale}
\end{equation}
for a 2D system. Here it should again be noted that $\chi_u$ vanishes for finite $L$ when $\beta \to \infty$, which implies that in this case ground state
results are useless. Instead, the above equation applies for results obtained with $\beta \propto L^z$. As demonstrated in Fig.~\ref{converg2}, for $\beta=2L$
the susceptibility is still clearly non-vanishing and we can attempt $z=1$ scaling. In Fig.~\ref{critical}, panel (a), we graph $L\chi_{u}$ in the region
$g\in[1.96,2.00]$ for different $L$, and for comparison also replot the stiffness data of Fig.~\ref{cross} for $g$ within this smaller window. Some drifts of the
crossing points can be seen in both quantities, more so in $L\chi_{u}$ than in $L\rho_{s}$.

\begin{figure}
\centering
{\resizebox*{0.5\textwidth}{!}{\includegraphics{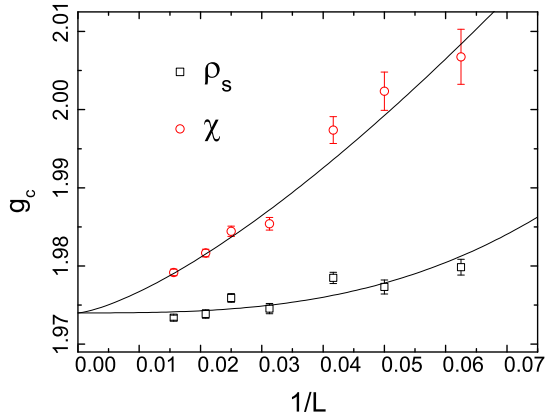}}}
\caption{(Color online). Size-dependent critical couplings $g_{c}(L)$ defined using $(L/2,L)$ crossing points of the spin stiffness and the uniform susceptibility
extracted from the data in Fig.~\ref{critical}. The two data sets were fitted to the form (\ref{criticalscale}), with different subleading exponents
($\omega \approx$ 3 and $\approx 2$ for the $\chi$ and $\rho_s$ fits, respectively) but constrained to a common critical point in the thermodynamic limit.
This fit resulted in the infinite-size critical point $g_c=1.974(4)$.
\label{criticalpoint}}
\end{figure}

To analyze the drifts quantitatively, we extract crossing points for pairs of systems of size, $L/2$ and $L$. To this end we fit polynomials of suitable
order to the data points and use them to extract the crossing point. The procedure was repeated $1000$ times with added Gaussian noise (with standard
deviation equal to the SSE error bars) in order to obtain a the statistical errors for the crossing points. The results for the so obtained finite-size estimates
(crossing points) $g_c(L)$ are graphed versus $1/L$ in Fig.~\ref{criticalpoint}. One in general expects the size corrections to be of the form
\begin{equation}
      g_{c}(L)=g_{c}(\infty)+aL^{-\omega},
\label{criticalscale}
\end{equation}
where $\omega$ is the sum of $1/\nu$ and a correction exponent. Since the correction in practice is an effective correction containing also higher order
contributions we let $\omega$ be different for the two quantities analyzed but the infinite-size critical point $g_c(\infty)$ is constrained to be the same
in both cases. The fitted functions are also shown in Fig.~\ref{criticalpoint}. Estimating the statistical fluctuations of the fits using Gaussian noise
as above, we finally obtain the critical point value  $g_{c}(\infty)=1.974(4)$ for our model with $p=1/2$.

\begin{figure}
\centering
{\resizebox*{0.5\textwidth}{!}{\includegraphics{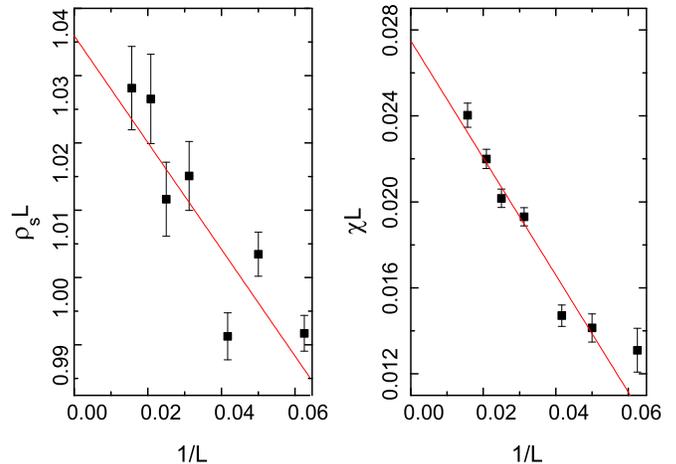}}}
\caption{(Color online). Size dependence of the vertical crossing points in Fig.~\ref{critical} for system-size pairs ($L/2,L$). The values based on
$L\rho_s$ and $L\chi_u$ are shown in the left and right panels, respectively, along with line fits.}
\label{criticaly}
\end{figure}

Next we analyze the vertical drifts of the crossing points. As shown in Fig.~\ref{criticaly} the behavior is consistent with a correction linear in
$1/L$, and, thus, there is a well defined crossing point for $L \to \infty$. This suggests that in fact the dynamic exponent really is $z=1$ as
we assumed in the preceding analysis.

\subsection{Quantum critical scaling at $T>0$}

To test the value of $z$ further, we carry out a scaling study of the uniform susceptibility in a completely different regime, with simulation data at elevated
temperatures and taking the limit $L \to \infty$. In the thermodynamic limit, in the neighborhood of the critical point, the susceptibility should
follow the form \cite{Yejingwu}
 \begin{equation}
        \chi_{u}= a+bT^{(2/z-1)},
\label{exponentz}
\end{equation}
for $T$ above a cross-over temperature when $g \not= g_c$ (and with $a<0$ for $g>g_c$ and $a>0$ for $g>g_c$). Exactly at the critical point $a=0$ and
the scaling behavior extends to $T=0$. The constant $b$ is non-universal, depending on a model-specific velocity. In principle this scaling form allows also
for an independent determination of the critical point, in addition to extracting $z$, by locating a $g$-value and exponent $z$ for which
$\chi_{u} \propto T^{(2/z-1)}$ holds. Here we just use the $g_c$ value obtained below and test the scaling with $z=1$.

\begin{figure}
\centering
{\resizebox*{0.5\textwidth}{!}{\includegraphics{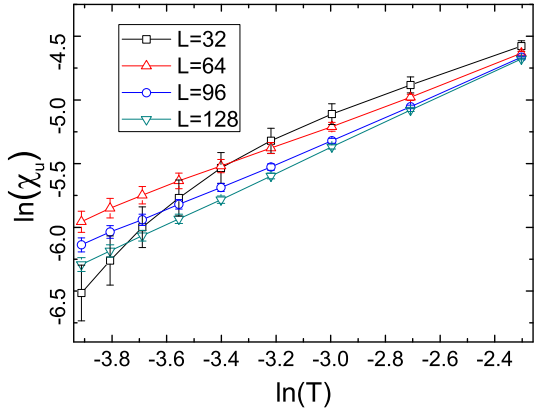}}}
\caption{(Color online). Uniform magnetic susceptibility versus temperature at $g=1.98$, for convenience graphed on a log-log scale. The
size dependence is non-monotonic and for $L=128$ only small finite-size effects should remain for the temperatures, $T \ge 0.02$ considered here.}
\label{sizeeffect}
\end{figure}

First, we check the convergence to the thermodynamic limit. Fig.~\ref{sizeeffect} shows data sets for different system sizes versus temperature. The size
dependence for fixed $T$ is non-monotonic at low $T$. For sufficiently large $L$ we expect exponential convergence. As shown in Fig.~\ref{sizeeffect}, for the
largest system studied, $L=128$, we have almost achieved convergence in the temperature range $T > 0.02$. The remaining size effects should be at most the size
of a few error bars, which is small on the scale of the overall variations with temperature. We therefore do not expect any significant distortions
by using these data instead of fully converged data.

\begin{figure}
\centering
{\resizebox*{0.5\textwidth}{!}{\includegraphics{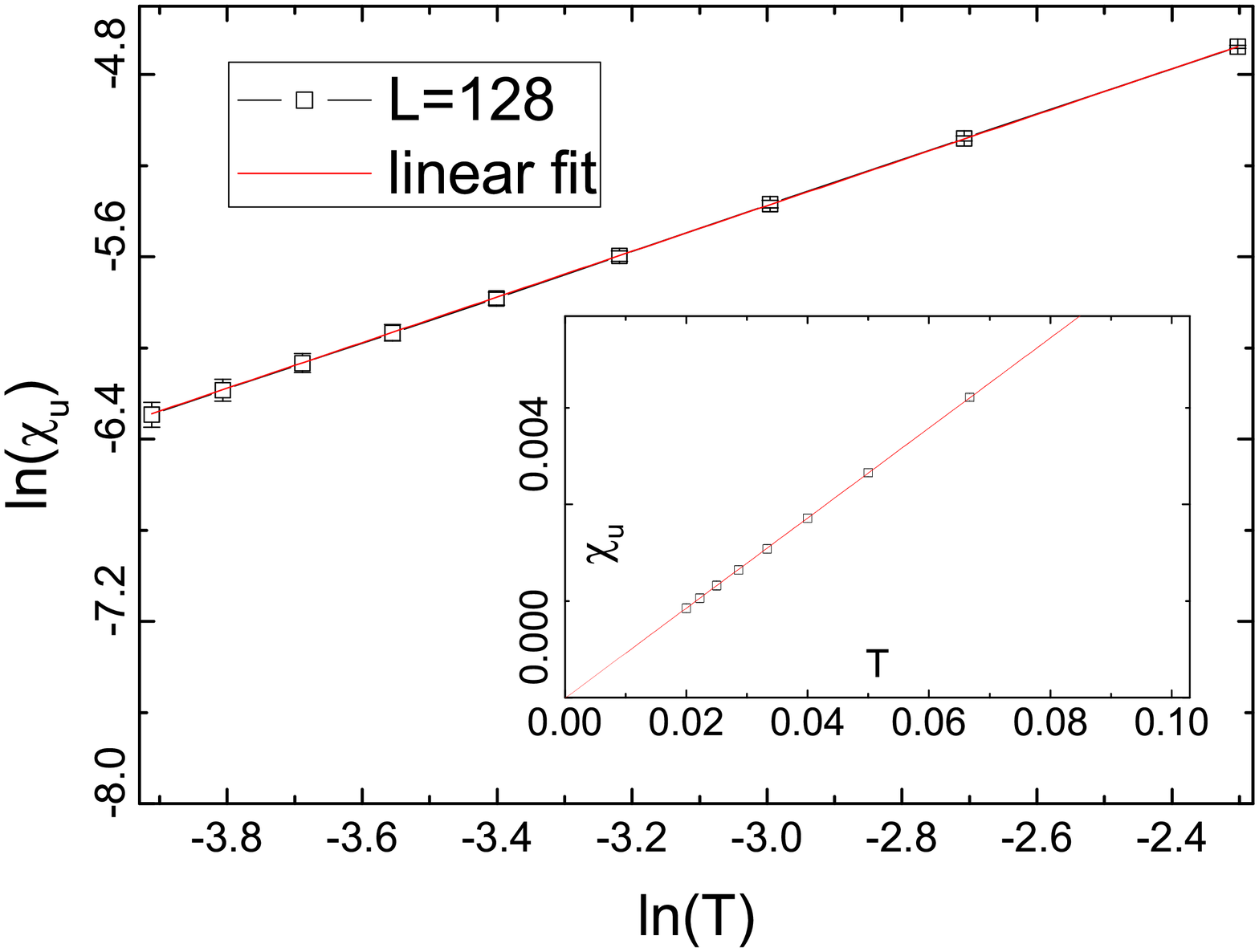}}}
\caption{(Color online). Temperature dependence of the uniform susceptibility at the estimated critical coupling, $g=1.974$, for $L=128$. The main panel
and the inset show the data on log-log and lin-lin scales, respectively. The data follow very closely the expected quantum-critical linear-$T$ form
(\ref{exponentz}) with $a=0$, $z=1$, and $b=0.093$, as shown with the straight lines.}
\label{zexponent}
\end{figure}

We plot the $L=128$ data both on log-log and lin-lin scales in Fig.~\ref{zexponent}. The behavior shows a remarkable consistency with $z=1$
critical scaling. A linear fit to the log-log data gives $z=1.00(1)$. Moreover, on the lin-lin plot it can be seen that the intercept $a$
vanishes (within the error bars), thus also independently confirming the location of the critical point determined above.

Our conclusion is, thus, that $z$ remains unchanged at unity when disorder is introduced in the system. This is in contrast to dimer-diluted systems, where
different dynamic exponents $z>1$ have been found.~\cite{sandvik02,vajk02,fulfill} An unchanged dynamic exponent was also found in dimerized models where the
coupling strengths are kept fixed but the configuration of dimers is random.\cite{yao10} These results are intriguing as one would normally expect disorder
introduced only in the space dimension to destroy the emergent Lorenz invariance (space-time symmetry implying $z=1$) present in the clean system.

\subsection{Critical exponents}

In order to study the disorder effects on the phase transition more completely, we also extract the standard static critical exponents. We
analyze the behaviors of the staggered structure factor $S(\pi,\pi)$ and the staggered susceptibility $\chi(\pi,\pi)$, which are expected to
have the following leading size dependencies at the critical point:
\begin{eqnarray}
S(\pi,\pi) &\sim & L^{2-z-\eta}, \label{sscale1} \\
\chi(\pi,\pi) &\sim& L^{2-\eta}. \label{xscale1}
\end{eqnarray}
In order to improve the fits when small systems are included, it is useful to also add corrections to scaling.
We then use the following forms:
\begin{eqnarray}
S(\pi,\pi) & = & aL^{2-z-\eta} + bL^{\omega_s}, \label{sscale2} \\
\chi(\pi,\pi) &=& cL^{2-\eta} + dL^{\omega_\chi}, \label{xscale2}
\end{eqnarray}
where of course the subleading exponents $\omega_S$ and $\omega_\chi$ should be smaller than the leading exponents.

\begin{figure}
\centering
{\resizebox*{0.5\textwidth}{!}{\includegraphics{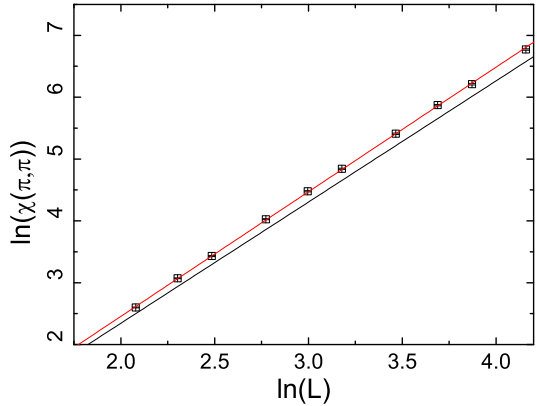}}}
\caption{(Color online). Dependence of the staggered susceptibility on the system size $L$ for $g=1.974$ (the estimated critical point). The linear
fit to the data (red line) is unsatisfactory when all system sizes are included, with $\chi^2/{\rm dof} \approx 4$. The slope is
$2-\eta = 2.015$. The black line shows the slope $\eta = 1.9625$ expected asymptotically in the standard O(3) universality class applicable
to the clean system.\cite{compostrini02}}
\label{linear}
\end{figure}

\begin{figure}
\centering
{\resizebox*{0.5\textwidth}{!}{\includegraphics{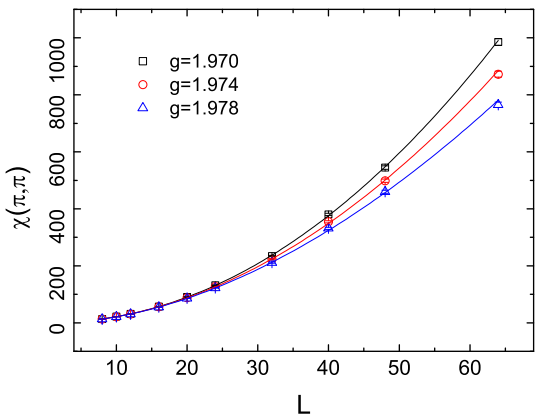}}}
\caption{(Color online). Finite-size scaling of the staggered susceptibility at the critical point. The data are the same as in Fig.~\ref{linear},
but here the fitted curve is of the form (\ref{xscale2}) including a correction to scaling. The leading power is now $2-\eta = 1.971(6)$, which is
consistent with the O(3) universality class, and the fit has $\xi^2/{\rm dof} \approx 1$. Going away from the estimated critical point by one error
bar, the exponents change to $2-\eta=2.046(9)$ for $g=1.970$ and $2-\eta=1.966(4)$ for $g=1.978$ ($\chi^2/{\rm dof}$ remains close to $1$ in
these fits).}
\label{scalex}
\end{figure}

The critical susceptibility analyzed only with the leading form (\ref{xscale1}) is shown on a log-log plot in Fig.~\ref{linear}. An essentially linear behavior
is seen, with the slope $2-\eta \approx 2.015$, i.e., $\eta \approx -0.015$. Without disorder, the system is in the standard O(3) universality class, where
$\eta =0.0375(15)$.\cite{compostrini02} While the fit visually looks reasonably good, the value of $\chi^2$ per degree of freedom is close to $4$, indicating
an only marginally satisfactory fit. The quality of the fit can be improved if some of the smaller system sizes are excluded, but then the statistical
accuracy of $\eta$ deteriorates. When we instead include the subleading correction, as shown in Fig.~\ref{scalex}, the fit improves significantly, with
$\chi^2/{\rm dof} \approx 1$. Interestingly, the exponent then also changes to $\eta =0.029(6)$, which is completely consistent with the above cited
best estimate of this exponent in the O(3) university class. For the subleading exponent we obtain $\omega_\chi = 0.5(4)$, further confirming that
the correction is small relative to the leading term.

The above estimate of $\eta$ does not take into account the uncertainty in the critical point. By repeating the fit including the scaling
correction at one error bar away from the mean value $g_c=1.974$, i.e., at $g=1.970$ and $1.978$, the leading exponent $2-\eta$ changes substantially,
as also seen in Fig.~\ref{scalex}. Based on the susceptibility we can then only conclude that $\eta$ is roughly within the range $[-0.05,0.04]$.

\begin{figure}
\centering
{\resizebox*{0.5\textwidth}{!}{\includegraphics{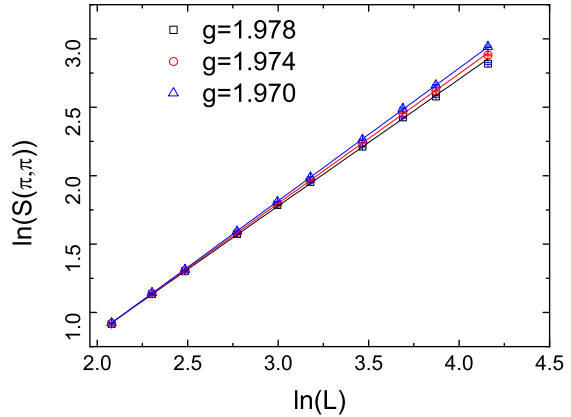}}}
\caption{(Color online). Scaling of the staggered structure factor at $g=1.970$, $1.974$, and $1.978$. Pure power-law fits
(lines on the log-log scale) give $1-\eta=0.927(5),0.948(1)$ and $0.970(1)$ with $\chi^2/{\rm dof} < 1$ in all cases.}
\label{scalesf}
\end{figure}

Moving now to the staggered structure factor, when plotting it on a log-log scale the slope should be $k=2-z-\eta$ according to Eq.~(\ref{sscale1}).
Using the O(3) value of $\eta$ and $z=1$ we should have $k \approx 0.962$. We present our SSE data along with the best-fit line on a log-log scale
at $g=1.974$ in Fig.~\ref{scalesf}. The slope is $k=0.948(1)$, a bit smaller than expected, and $\chi^2/{\rm dof} < 1$. In this case the subleading
correction does not improve the fit statistically, but this of course does not exclude that there are some effects of corrections. Repeating the
fit at $g_c$ plus and minus one error bar again introduces a fluctuation in the slope, and based on all these fits (with the assumption that $z=1$)
we have $\eta=0.05(2)$, in excellent agreement with the best $\eta$ values calculated for the O(3) class. Note that the primary reason why the
statistical accuracy of $\eta$ is better when extracted from fits to $S(\pi,\pi)$ than $\chi(\pi,\pi)$ is that no correction to scaling is required
in the former case.

The second important static exponent is the correlation length exponent $\nu$, which governs the divergence of the correlation length and is accessible
in finite-size scaling through the argument of scaling functions such as Eq.~\ref{fsscform}. This exponent is at the heart of the Harris criterion, according
to which a unique transition point is possible in the presence of disorder only if the exponent satisfies $\nu\geq 2/d$, where in quantum systems it is
believed that $d$ should be the spatial dimensionality (with there being no disorder in the time direction in the path integral representation). Thus,
the introduction of disorder in a clean system with $\nu < 2/d$ should lead to a new universality class in which the Harris inequality is satisfied.
We have shown above that the exponents $z$ and $\eta$ in the $J_1-J_2-J_3$ model are not changed by the disorder and we now examine $\nu$.

\begin{figure}
\centering
{\resizebox*{0.5\textwidth}{!}{\includegraphics{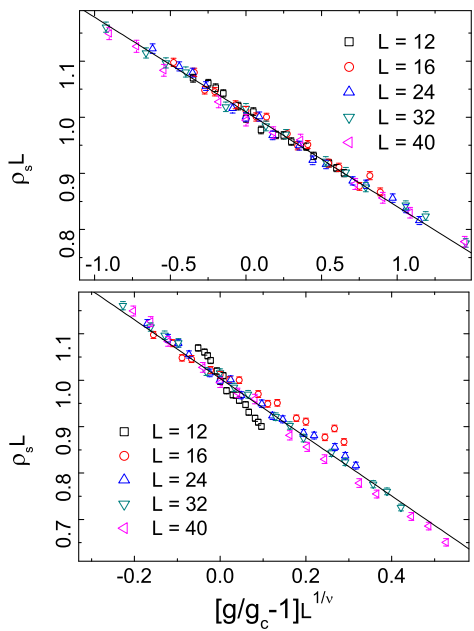}}}
\caption{(Color online). Tests of scaling collapse of the spin stiffness, using the finite-size scaling form (\ref{fsscform}) with SSE
data at $g=1.974$. Upper panel: With the correlation-length exponent at its O(3) value\cite{compostrini02} $\nu=0.7115$. The data collapse
is good, with $\chi^2/{\rm dof} = 1.02$ computed relative to a fitted polynomial. Lower panel, using $\nu=1$, corresponding to the lower
edge of the range of values satisfying the Harris inequality $\nu \ge 2/d$. Here the data collapse is visibly much less satisfactor,
and quantitatively the goodness is $\chi^2/{\rm dof} \approx 6.5$.}
\label{nu}
\end{figure}

Based on the scaling function (\ref{fsscform}) for the spin stiffness and assuming $z=1$, plotting $\rho_sL$ versus $(g-g_c)L^{1/\nu}$, curves for different
$L$ should collapse onto a common scaling function for large $L$. Based on the already extracted exponents $z$ and $\eta$ it seems plausible that the
universality class remains the 3D O(3) class, and we therefore test data collapse with the corresponding value of $\nu$, for which the best estimate to
our knowledge is from Ref.~\onlinecite{compostrini02}, $\nu = 0.7115()$. The result of this procedure is a statistically good collapse of the data, as
illustrated in the upper panel of Fig.~\ref{nu}. To quantify the goodness of the collapse, we fit a polynomial to all the data points, and this has a
satisfactory value of $\chi^2/{\rm dof} \approx 1$. If the Harris inequality hods, the smallest possible value of the exponent is $\nu=1$. The lower
panel of Fig.~\ref{nu} shows the result of an attempted data collapse with this value. While some of the data do fall close to a common curve, there
are also segments of points that deviate very clearly, and this is quantified with a statistically unsatisfactory goodness of a fit of a
polynomial, which has $\chi^2/{\rm dof} > 6$. Thus, while our data do not allow a very precise estimate of $\nu$, we can clearly see that the
Harris criterion is violated.

\section{Susceptibility of the Mott Glass}
\label{sec:mg}

In the previous section we studied the point at which the antiferromagnetic long-range order vanishes in a system with
random coupling constants. In disordered systems one in general expects a Griffiths phase following this kind of critical point. In the
Griffiths, or glassy (here quantum glassy) phase there are arbitrarily large clusters of the ordered phase inside a background of the disordered
phase, leading to singular behaviors. The critical point extracted in the previous section is that separating the quantum
glass and the Ne\'el state. The boundary between the glass state and the eventual quantum paramagnetic state is not easy to determine, because
the Griffiths singularities are due to large clusters, which are very rare close to the phase boundary.
We here study the finite-temperature behavior of the uniform magnetic susceptibility inside the glass phase and show that it has a particular
behavior that in principle should be useful for also detecting the glass--paramagnetic phase boundary.

In the quantum paramagnetic phase the gapped magnons in combination with the 2D density of triplets leads to a pure exponential form of
the susceptibility,
 \begin{equation}
     \chi\sim\exp(-\Delta /T),
\label{parasusc}
\end{equation}
where $\Delta$ is the singlet-triplet gap. As we discussed in Sec.~\ref{sec:intro}, the glass phase of a spin-isotropic Heisenberg system is expected to be
of the Mott type, i.e., gapless but with the susceptibility still vanishing as $T \to 0$. To our knowledge, the only quantitative study of the form of
the compressibility in a 2D spin system is in Ref.~\onlinecite{yu05}, where in a diluted $S=1$ model the stretched exponential form
 \begin{equation}
     \chi\sim\exp(-b/T^{\alpha}),
\label{suscrelation}
\end{equation}
was found, with $\alpha=1/2$. No theoretical explanation of this form was provided. In Ref.~\onlinecite{wang11}, in a study of the compressibility of a disordered
Boson system, the same general form was found but with $\alpha$ varying ($\alpha < 1$), and it was argued that this form follows naturally from a scenario where
there is a distribution of cluster sizes of the non-gapped phase inside the gapped phase, with the probability of a site depending to a cluster of a given
size $s$ decaying exponentially with $s$ (as in percolation theory below the percolation threshold), and if the finite-size gaps of those
clusters depend as a power-law on their size. For the random dimerized Heisenberg system we here also find the stretched exponential
form with varying $\alpha$, and it seems very plausible that this form here is also due to the same conditions as mentioned above
(with finite-size gapped N\'eel-phase clusters in a background of the dominant quantum-paramagnetic phase).

\begin{figure}
\centerline{\includegraphics[angle=0,width=8.5cm, clip]{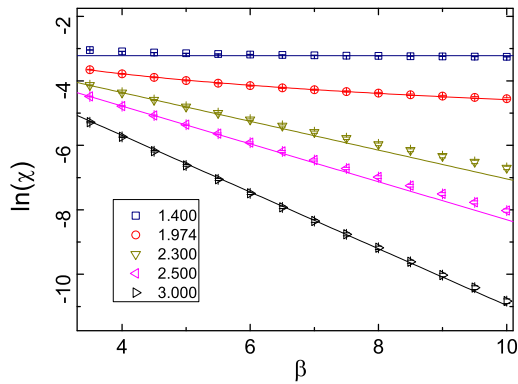}}
\caption{(Color online). Uniform susceptibility vs inverse temperature for $L=32$ systems. When $g<1.974$ (in the N\'eel phase), the curves
flatten out and converge to a non-zero constant (flat line). For $g=g_c \approx 1.974$ the $T$-linear behavior applies, as shown with the fitted curve.
When $g=3$ a standard exponential decay ${\rm e}^{-\Delta/T}$ is found (fitted curve shown), indicating a gapped quantum paramagnet, while for smaller
$g > g_c$ there are significant deviations from the exponential form (the lines are here drawn to go through the data for the higher temperatures), suggesting
the system is a Mott glass.}
\label{glassphasetrans}
\end{figure}

\begin{figure}

\centering
{\resizebox*{0.5\textwidth}{!}{\includegraphics{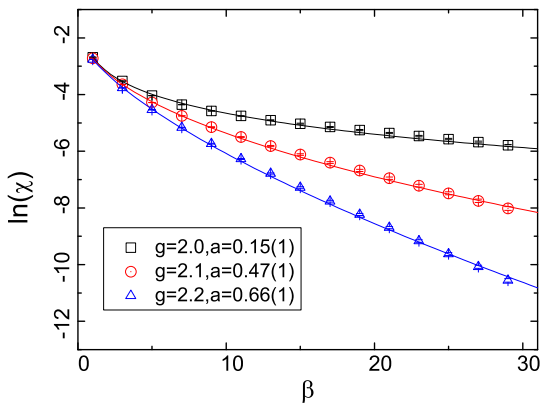}}}
\caption{(Color online). Uniform susceptibility of systems in the MG phase. The fitted curves are of the form (\ref{suscrelation})
with $\alpha=0.15(1),0.47(1),0.66(1)$ for $g=2.0,2.1$, and $2.2$, respectively.}
\label{glassphase}
\end{figure}

To study the $T$ dependent susceptibility we again first investigated finite size effects and in results presented here go down only to temperatures where we
can avoid them. When the length of the system is larger than $L=32$ finite-size effects can be excluded down to $T=0.1$. The temperature dependence of
$\ln(\chi_{u})$ is shown in Fig.~\ref{glassphasetrans} for this system size and different $g$ values. When $g \gg g_c$, deep inside the quantum paramagnetic phase, the
standard exponential form (\ref{suscrelation}) applies, while in the N\'eel phase the temperature dependence is very weak at low $T$. Between these
well-understood regimes we find a clearly decreasing susceptibility as $T \to 0$, but not with the pure exponential form (\ref{parasusc}). As shown in
detail for three $g$-values in Fig.~\ref{glassphase}, we instead see the stretched exponential form (\ref{suscrelation}) with the exponent $\alpha$ depending
on $g$. This form indicates that the system is gapless but there is no magnetic order (in which case the susceptibility must be non-zero for $T\to 0$ by
hydrodynamic arguments \cite{nelson}). The stretched exponential should then be a characteristic of the MG state. It is not easy to locate the phase
boundary between the MG to quantum paramagnet, but in principle a detailed study of the exponent $\alpha$ versus $g$ should allow for this.


\section{summary and discussion}
\label{sec:disc}

We have introduced a dimerized 2D antiferromagnetic Heisenberg model with three coupling constants $J_1 < J_2 < J_3$, where the $J_2$ and $J_3$
couplings reside randomly on bonds forming columns on the simple square lattice. We have studied the ground state phases and quantum phase transitions
of the model using unbiased SSE QMC simulations on lattices with several thousand spins. Using finite-size and finite-temperature scaling of several
different quantities, we have shown that, in spite of the presence of relatively strong disorder (the ratio of the two intra-dimer couplings being
$\approx 1.6$ at the critical point), the universality class of the system remains in the same standard O(3) universality class as clean dimerized
Heisenberg models. The Harris criterion for relevance of disorder is therefore violated.

Violation of the Harris criterion
was previously found in a dimer system with no randomness in the values of intra- and inter-dimer couplings, but with the dimers
arranged randomly.\cite{yao10} Evidence is this mounting for very generic non-applicability of the Harris criterion in a broad
class of dimerized Heisenberg models. To our knowledge, the only case where the Harris criterion was shown to be valid in dimerized quantum
spin models was for a bilayer where some fraction of dimers were completely removed in a random fashion.\cite{sandvik02,vajk02,fulfill,sandvik06}
Naively one might not expect these different cases of disorder to lead to different universality classes (since the symmetries are the same)
and the reasons why they are different certainly deserve further examination. Given the loopholes \cite{pazmandi97} of the Harris original \cite{harris}
and generalized \cite{chayes86} analysis of critical points in the presence of disorder, it is now becoming clear that the issue of relevance
or irrelevance is much more complex than previously believed.

We also found a Mott glass phase by tuning the disorder and interaction strength beyond the critical point where the N\'eel order vanishes. Here
the system is incompressible, as in the quantum paramagnet reached eventually, but there is no magnetic order and the susceptibility follows a stretched
exponential form, $\chi\sim\exp(-b/T^{\alpha})$ with $0<\alpha<1$, instead of the normal exponential form dictated by the gap in the quantum paramagnetic
phase (i.e., $\alpha=1$ and $b$ equals the singlet-triplet gap). The natural interpretation of the stretched exponential is that, within the Griffiths scenario
of domains of N\'eel-phase clusters inside the quantum paramagnet, the clusters have finite size gaps decaying as a power-law with the size of the clusters
(as has been argued also in the context of a disordered boson model.\cite{wang11}. In a previous study of a disordered $S=1$ system the same kind of
stretched exponential form was found,\cite{yu05} but in that case $\alpha=1/2$ was always observed. In our case the exponent is clearly varying, as
was found also in the glass phase of disordered bosons.\cite{wang11}

\begin{acknowledgments}
We would like to thank Wenan Guo for stimulating discussions.
NSM and DXY acknowledge support from National Basic Research Program of China (2012CB821400), NSFC-11074310, NSFC-11275279, Specialized Research Fund for the
Doctoral Program of Higher Education (20110171110026), and NCET-11-0547. The work of AWS was supported by the NSF under Grants No.~DMR-1104708 and PHY-1211284,
and he also gratefully acknowledges support from Sun Yat-Sen University for visits during which part of this research was completed.
\end{acknowledgments}


\begin{thebibliography}{99}

\bibitem{overall}
M. Vojta, Rep. Prog. Phys. {\bf 66}, 2069 (2003).

\bibitem{lee85}
P. A. Lee and T. V. Ramakrishnan. Rev. Mod. Phys. {\bf 57}, 287 (1985).

\bibitem{abrahams01}
E. Abrahams, S. V. Kravchenko, and M. P. Sarachik, Rev. Mod. Phys. {\bf 73}, 251 (2001).

\bibitem{fisher}
M. p. A. Fisher, P. B. Weichman, G. Grinstein, and D. S. Fisher, Phys. Rev. B {\bf 40}, 546 (1989).

\bibitem{BEC}
J. E. Lye, L. Fallani, M. Modugno, D. S. Wiersma, C. Fort, and M. Inguscio, Phys. Rev. Lett. {\bf 95}, 070401 (2005).

\bibitem{hefour}
G. A. Cs\'athy, J. D. Reppy, and M. H. W. Chan, Phys. Rev. Lett. {\bf 91}, 235301 (2003).

\bibitem{fisher94}
 D. S. Fisher, Phys. Rev. B {\bf 50}, 3799 (1994).

\bibitem{bbec}
R. Yu, C. F. Miclea, F. Weickert, R. Movshovich, A. Paduan-Filho, V. S. Zapf, and T. Roscilde, Phys. Rev. B {\bf 86}, 134421 (2012).

\bibitem{nature12}
R. Yu, L. Yin, N. S. Sullivan, J. S. Xia, C. Huan, A. Paduan-Filho, N. F. Oliveira Jr, S. Haas, A. Steppke, C. F. Miclea, F. Weickert, R. Movshovich, E.-D. Mun, B. L. Scott,
V. S. Zapf, T. Roscilde, and A. Kitaev, Nature {\bf 489}, 379 (2012).

\bibitem{superconductor}
Q. Luo, D. X. Yao, A. Mereo, and E. Dagotto, Phys. Rev. B {\bf 83}, 174513 (2011).

\bibitem{wangperc}
L. Wang and A. W. Sandvik, Phys. Rev. Lett. {\bf 97}, 117204 (2006); Phys. Rev. B {\bf 81}, 054417 (2010).

\bibitem{changlani13}
H. J. Changlani, S. Ghosh, S. Pujari, and C. L. Henley, Phys. Rev. Lett. {\bf 111}, 157201 (2013).

\bibitem{basko06}
D. Basko, I. Aleiner, and B. Altshuler, Ann. Phys. (NY) {\bf 321}, 1126 (2006).

\bibitem{sachdev}
S. Sachdev,  {\it Quantum Phase Transitions}, (Cambridge University Press, 1999).

\bibitem{manousakis91}
E. Manousakis, Rev. Mod. Phys. {\bf 63}, 1 (1991).

\bibitem{sandvik10}
A. W. Sandvik, AIP Conf. Proc. {\bf 1297}, 135 (2010).

\bibitem{chen00}
Y.-C. Chen and A. H. Castro Neto, Phys. Rev. B {\bf 61}, R3772 (2000).

\bibitem{kato00}
K. Kato, S. Todo, K. Harada, N. Kawashima, S. Miyashita, and H. Takayama, Phys. Rev. Lett. 84, 4204 (2000)

\bibitem{sandvik01}
A. W. Sandvik, Phys. Rev. Lett. {\bf 86}, 3209 (2001).

\bibitem{sandvik02}
A. W. Sandvik, Phys. Rev. B 66, 024418 (2002).

\bibitem{yu05}
R. Yu, T. Roscilde, and S. Haas, Phys. Rev. Lett. 94, 197204 (2005); Phys. Rev. B 73, 064406 (2006)

\bibitem{laflorencie06}
N. Laflorencie, S. Wessel, A. L\"auchli, and H. Rieger, Phys. Rev. B {\bf 73}, 060403(R) (2006).

\bibitem{liu06}
C.-W. Liu, S. Liu, Y.-J. Kao, A. L. Chernyshev, and A. W. Sandvik,
Phys. Rev. Lett. 102, 167201 (2009).

\bibitem{vajk02}
O. Vajk, P. Mang, M. Greven, P. Gehring, and J. Lynn, Science {\bf 295}, 1691 (2002).

\bibitem{carretta11}
P. Carretta, G. Prando, S. Sanna, R. De Renzi, C. Decorse, and P. Berthet,
Phys. Rev. B {\bf 83}, 180411(R) (2011).

\bibitem{glassphase}
N. Prokof'ev and B. Svistunov, Phys. Rev. Lett. {\bf 92}, 015703 (2004).

\bibitem{giamarchi01}
T. Giarmarchi, P. Le Doussal, and E. Orignac, Phys. Rev. Lett. {\bf 64}, 245119 (2001).

\bibitem{boson1d}
E. Altman, Y. Kafri, A. Polkovnikov, and G. Refael, Phys. Rev. Lett. {\bf 93}, 150402 (2004).

\bibitem{wang11}
Y. Wang, W. Guo, and A. W. Sandvik, arXiv:1110.3213.

\bibitem{mglass03}
S. Iyer, D. Pekker, and G. Refael, Phys. Rev. B {\bf 85}, 094202 (2012).

\bibitem{yao10}
D. X. Yao, J. Gustafsson, E. W. Carlson, and A. W. Sandvik, Phys. Rev. B {\bf 82}, 172409 (2010).

\bibitem{roscilde07}
T. Roscilde, and S. Haas,  Phys. Rev. Lett. {\bf 99}, 047205 (2007).

\bibitem{harris}
A. B. Harris, J. Phys. C {\bf 7}, 1671 (1974).

\bibitem{chayes86}
J. T. Chayes, L. Chayes, D. S. Fisher, and T. Spencer, Phys. Rev. Lett. {\bf 57}, 2999 (1986).

\bibitem{singh88}
R. R. P. Singh, M. P. Gelfand, and D. A. Huse, Phys. Rev. Lett. 61, 2484 (1988).

\bibitem{matsumoto01}
M. Matsumoto, C. Yasuda, S. Todo, and H. Takayama, Phys. Rev. B 65, 014407 (2001).

\bibitem{wenzel09}
S. Wenzel. and W. Janke, Phys. Rev. B 79, 014410 (2009).

\bibitem{sse}
A. W. Sandvik, Phys. Rev. B {\bf 56}, 11678 (1997), {\it ibid} {\bf 59}, 14157 (1999).

\bibitem{fulfill}
R. Sknepnek, T. Vojta, and M. Vojta, Phys. Rev. Lett. {\bf 93}, 097201 (2004).

\bibitem{kisker97}
J. Kisker and H. Rieger, Phys. Rev. B {\bf 55}, R11981 (1997).

\bibitem{pazmandi98}
F. P$\acute{a}$zm$\acute{a}$ndi and G. T. Zim$\acute{a}$ny, Phys. Rev. B {\bf 57} 5044 (1998).

\bibitem{pazmandi97}
F. P$\acute{a}$zm$\acute{a}$ndi, R. T. Scalettar, and G. T. Zim$\acute{a}$ny, Phys. Rev. Lett. 79, 5130 (1997).

\bibitem{classh}
T. Vojta, and R. Sknepnek, Phys. Rev. B {\bf 74}, 094415 (2006).

\bibitem{qmch}
A. W. Sandvik, Phys. Rev. Lett. {\bf 96}, 207201 (2006).

\bibitem{nelson}
S. Chakravarty, B. I. Halperin, and D. R. Nelson, Phys. Rev. Lett. {\bf 60}, 1057 (1988).

\bibitem{sse2}
A. W. Sandvik, Phys. Rev. B {\bf 56}, 11678 (1997).

\bibitem{conver1}
S. Wenzel, and W. Janke, Phys. Rev. B {\bf 79}, 014410 (2009).

\bibitem{conver2}
W. Krauth, N. Trivedi, and D. Ceperley, Phys. Rev. Lett. {\bf 67}, 2307 (1991).

\bibitem{cldimer}
S. Wenzel, L. Bogacz, and W. Janke, Phys. Rev. Lett. {\bf 101}, 127202 (2008).

\bibitem{Yejingwu}
A. V. Chubukov, S. Sachdev, and J. Ye, Phys. Rev. B {\bf 49}, 11919 (1994).

\bibitem{offdiagonal}
P. Sengupta, and S. Haas, Phys. Rev. Lett. {\bf 99}, 050403 (2007).

\bibitem{sandvik02b}
A. W. Sandvik, Phys. Rev. Lett. {\bf 89}, 177201 (2002)

\bibitem{vajk02b}
O. P. Vajk and M. Greven, Phys. Rev. Lett. {\bf 89}, 177202 (2002)

\bibitem{sandvik06}
A. W. Sandvik, Phys. Rev. Lett. 96, 207201 (2006)

\bibitem{compostrini02}
M. Campostrini, M. Hasenbusch, A. Pelissetto, P. Rossi, and E. Vicari, Phys. Rev. B {\bf 65}, 144520 (2002).

\end{thebibliography}
\end{document}